\documentclass[letter]{aa}
\usepackage{natbib}
\bibpunct{(}{)}{;}{a}{}{,} 
\usepackage{graphicx}
\usepackage{txfonts}
\usepackage{mathtools}
\usepackage[colorlinks=true,linkcolor=blue,urlcolor=blue,citecolor=blue]{hyperref}
\makeatletter
\renewcommand*\aa@pageof{, page \thepage{} of \pageref*{LastPage}}
\makeatother

\usepackage{xcolor}

\newcommand{\MH}{\ensuremath{\left[\mathrm{M}/\mathrm{H}\right]}}
\newcommand{\alphafe}{\ensuremath{\left[\alpha/\mathrm{H}\right]}}

\begin{document} 

\title{The Gaia DR3 view of dynamical substructure in the stellar halo near the Sun}
\author{Emma Dodd \and
        Thomas M. Callingham,
        Amina Helmi, 
        Tadafumi Matsuno, 
        Tom\'{a}s Ruiz-Lara, 
         Eduardo Balbinot 
        \and Sofie L\"{o}vdal 
          }

 \institute{Kapteyn Astronomical Institute, University of Groningen, Landleven 12, 9747 AD Groningen, The Netherlands\\
              \email{dodd@astro.rug.nl}
}

 \date{Received XXXX; accepted yyyy}

  \abstract
   {The debris from past merger events is expected and, to some extent, known to populate the stellar halo near the Sun. }
   {We aim to identify and characterise such merger debris using Gaia DR3 data supplemented by metallicity and chemical abundance information from LAMOST LRS and APOGEE for halo stars within 2.5 kpc from the Sun.}
   {We utilise a single linkage-based clustering algorithm to identify over-densities in Integrals of Motion space that could be due to merger debris. Combined with metallicity information and chemical abundances, we characterise these statistically significant over-densities.}
   {We find that the local stellar halo contains 7 main dynamical groups, some of in-situ and some of accreted origin, most of which are already known. We report the discovery of a new substructure, which we name ED-1. 
   In addition, we find evidence for 11 independent smaller clumps, 5 of which are new: ED-2, 3, 4, 5 and 6 are typically rather tight dynamically, depict a small range of metallicities, and their abundances when available, as well as their location in Integrals of Motion space, suggest an accreted origin.}
   {The local halo contains an important amount of substructure, of both  in-situ and accreted origin. }


   \maketitle
%

\section{Introduction}

The \textit{Gaia} mission has brought our Galaxy into sharper focus with every data release,
revolutionising our understanding of our local environment and the field of Galactic Archaeology. 
Notably, the second data release \citep{Gaia-DR2} significantly increased the number of stars with full 6-d position and velocity information.
This increase brought insights into our Galaxy's past, such as evidence of an ancient major merger \citep[known as {\it Gaia}-Enceladus,][see also \citealt{Belokurov2018}, the ``Sausage'']{helmi_merger_2018},
and fine details of the dynamics of the Galactic disk \citep[e.g.][]{antoja_dynamically_2018}.
The recent third data release \citep{GaiaDR3_Hack_Summary},
promises to offer similar advancements in our understanding of the Galaxy. 

Over the Milky Way's history, many galaxies must have been accreted in a series of minor and major mergers,
following the  hierarchical growth characteristic of the $\Lambda$CDM model \citep{springel_simulations_2005}. Inferring our assembly history from the accreted material means overcoming the challenge of identifying the accreted stars and attributing these to their progenitor. For all but the most recent events, the material has long since phase mixed, erasing cohesion in physical space.
Instead, we may look to the space of integrals of motion (IoM), where some structure is preserved \citep[see][and references therein]{helmi_streams_20}.
Combined with chemical abundances this can help identify a star's progenitor. 

This goal is currently a large endeavour in the Galactic community, and many structures have been recently identified in the stellar halo. 
Some of the larger ones have been studied for decades and are well established, such as Sagittarius and the Helmi streams,  and some are more recent discoveries such as   {\it Gaia}-Enceladus/Sausage.
However, the existence and extent of some other structures is debated \citep[see e.g.][]{naidu_evidence_2020}.

As the available data improves and grows in size,
the methods used to identify substructures have become increasingly sophisticated.
Nonetheless, the interpretation and statistical soundness of the outcome have generally received less attention. 
With this in mind, in this Letter, we apply our previously developed clustering algorithm \citep{lovdal2022,ruizlara2022}
to identify merger debris and in-situ substructures in the new {\it Gaia}  DR3 dataset.
This work is organised as follows.
Section 2 describes our selection of a \textit{Gaia} DR3 halo sample, complemented with chemistry.
We describe our methodology in section 3, and present our results and give a brief discussion in section 4.
In section 5, we summarise our findings.

\section{Data}\label{sec:data}
\textit{Gaia} DR3 has provided a significant increase (roughly a factor 5) in the number of stars with a radial velocity
\citep[the RVS sample,][]{Katz2022}. 
Furthermore, \textit{Gaia} DR3 provides, for the first time, metallicities for over 5 million  
stars derived from the RVS spectra 
\citep{RecioBlanco2022}.
As we show below, the increase in size and content information of this dataset offers new insights into the local stellar halo. 

To construct a sample suitable for our purposes, we apply several quality and selection cuts to the RVS dataset.
We first correct each star's parallax ($\varpi$) by their individual zero-point offsets ($\Delta_\varpi$),
determined following \citet{lindegren2021}.
To obtain a distance we invert the parallax, and hence we require that the (total) relative parallax uncertainty is less than 20\%,
i.e.  $(\varpi -\Delta_\varpi) /\sqrt{\sigma_{\textrm{parallax}}^2 + \sigma_{\textrm{sys}}^2} \ge 5 $,
where $\sigma_{\textrm{parallax}}$ is \texttt{parallax\_error}, 
and $\sigma_{\textrm{sys}}$ is the systematic uncertainty on the zero-point, which we take to be 0.015 mas  \citep{lindegren2021}.
Furthermore, we select stars with \texttt{RUWE} $< 1.4$ and $\sigma(V_{\rm los}) < 20$~km/s,   after applying the correction to \texttt{radial\_velocity\_error} recommended by  \citet{Babusiaux2022}. We also follow these authors and remove a few stars with  $(G_{\textrm{RVS}} - G) < -3$.

To make a kinematic selection of the local halo, we derive the velocities of the stars after
correcting for the solar motion using 
$ (U, V, W )_\odot$ = (11.1, 12.24, 7.25) km/s \citep{schonrich2010}
and for the motion of the local standard of rest (LSR) using a $|\textbf{V}_{\textrm{LSR}}|$ of 232.8\,km/s \citep{mcmillan2017}, and 
require  |\textbf{V}-\textbf{V}$_\textrm{LSR}$| $>$ 210 km/s. We
adopt $R_{\odot}$ = 8.2\,kpc \citep{mcmillan2017} and impose a distance cut of 2.5 kpc.
For stars at low latitude ($|b| < 7.5$), we require higher SNR spectra (\verb|rv_expected_sig_to_noise|~$>5$) 
to avoid highly contaminated spectra and spurious velocities, following
\citet[][Sec.~9]{Katz2022}. The resulting sample has 69,106 nearby halo stars.

To complement the dynamical information, we consider several sources of stellar chemistry data. We recalibrate the Gaia DR3 GSP-SPEC \MH{} and \alphafe{} abundances according to the recipes given in \citet{RecioBlanco2022},
and  follow \citet{RecioBlanco2022b}  to define a ``Medium quality" sample. 
This yields 4665 stars in our sample with a (reliable) [M/H] measurement. 
Additionally, our sample contains 1809 stars in APOGEE DR17 \citep{accetta2022} and 9797 stars in LAMOST LRS DR7 \citep{zhao2012}, with 675 stars in common.

\section{Methods}\label{sec:method}

To identify accreted debris in the local halo,
we apply a clustering algorithm to the three-dimensional Integrals of Motion (IoM) space of energy, and the 
$z$ and perpendicular components of angular momentum ($E$, $L_z$, $L_{\perp}$).
We compute $E$ using the same potential as in  \citet{lovdal2022}.
This potential consists of a Miyamoto-Nagai disk with parameters $(a_d, b_d) = (6.5, 0.26)$ kpc, $M_{d}=9.3\times 10^{10} M_\odot$,
a Hernquist bulge with $c_b = 0.7$ kpc, $M_{b}=3.0 \times 10^{10} M_\odot$ 
and an NFW halo with  $r_s=21.5$ kpc, $c_h$=12, and $M_{\rm halo}=10^{12} M_\odot$.
We define 
$L_z$ to be positive for prograde stars,
while $L_{\perp}$ = $\sqrt{L_x^2 + L_y^2}$.
Whilst $L_{\perp}$ is not a true IoM, it is approximately conserved, 
and is therefore useful to identify  halo substructure. 
We require  that all stars are bound in this potential, resulting in a final nearby halo sample of 68,921 stars.

The clustering algorithm is described in detail in \citet{lovdal2022} and \citet{ruizlara2022}, where it was applied to a local halo sample from \textit{Gaia} EDR3. We refer the reader to those papers for more information. 
It is based on the single linkage algorithm, which, at each step,
joins together the two closest components until all components are linked.
To determine where to stop the linkage and identify significant components/clusters, we determine 
at each step, how over-dense a cluster is 
relative to a sample  of 1000 artificial, smooth datasets obtained by re-shuffling the velocities of the stars.
That is, we compare the number of stars, $N_{C_{i}}$, in an ellipsoidal region centred on each cluster,
to the number of stars in our artificially generated smooth halos within the same region, 
$N^{\mathrm{art}}_{C_{i}}$.
The significance is then:
$S = (N_{C_{i}} - \langle  N^{\mathrm{art}}_{C_{i}}\rangle) / \sigma_{i}$, 
where $\sigma_{i} = \sqrt{N_{C_{i}} + (\sigma^{\mathrm{art}}_{C_{i}})^2}$
Our final set of clusters are extracted at their maximum significance,
and we keep clusters with a significance $S >3$ and a minimum number of 10 members.

As demonstrated by \citet{ruizlara2022}, the clusters identified by the algorithm are not necessarily physically independent from each other,
and can potentially be grouped together to form larger structures. To this end, 
 we define the Mahalanobis distance between two clusters in IoM space as
\begin{equation}
    D^{\prime} = \sqrt{({\boldsymbol \mu_1}-{\boldsymbol \mu_2})^T (\Sigma_1+\Sigma_2)^{-1} ({\boldsymbol \mu_1}-{\boldsymbol \mu_2})}\label{eq:mah_distance_distributions}
\end{equation}
where $\boldsymbol{\mu}_1, \boldsymbol{\mu}_2$ and $\Sigma_1, \Sigma_2$ correspond to the means and covariance matrices of the two cluster (ellipsoidal)  distributions respectively. $D^{\prime}$ thus gives a relative measure of how close clusters are in IoM space.
This distance metric may be visualised in a dendrogram, and can thus be used for a second stage of linkage between clusters. By
introducing a preliminary distance cut we can identify larger groups as well as individual clusters, which we then proceed to characterise dynamically and chemically. 

\section{Results}\label{sec:results}

\begin{figure*}
    \centering
    \includegraphics[width=\linewidth]{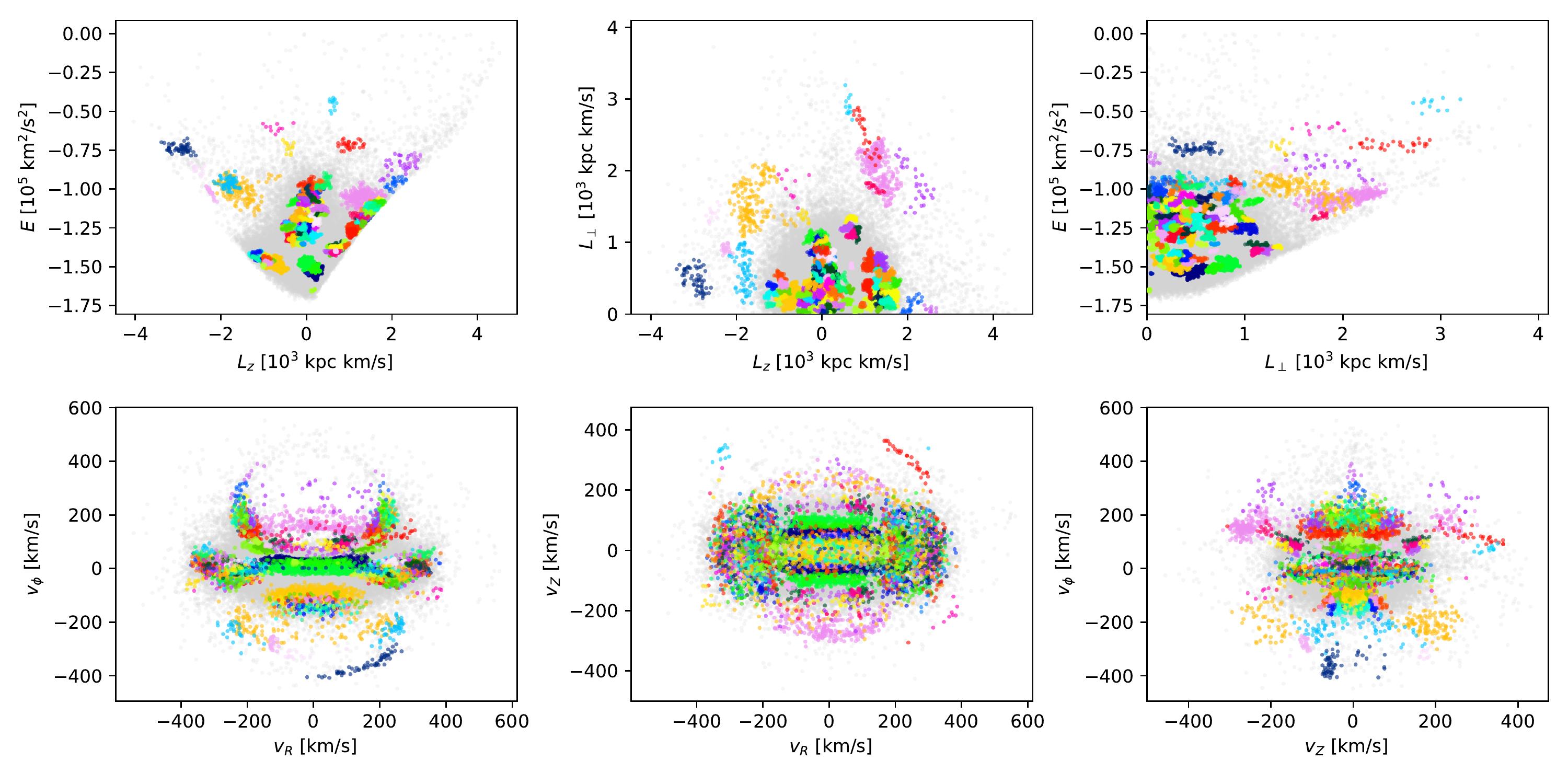}
    \caption{
    The members of the 91 significant clusters identified by our algorithm in our stellar halo sample,
    where different colours indicate a stars association with a cluster.
    Stars not attributed to a cluster are shown in the background in grey. 
    The  top and bottom rows show respectively, IoM and velocity space.
    The relation between the clusters is given as a dendrogram  in Fig. \ref{fig:dendrogram},
    and the joined groups in the same spaces in Fig. \ref{fig:IoM_Groups}.
    }
    \label{fig:IoM_Clusters}
\end{figure*}

Fig.~\ref{fig:IoM_Clusters} shows the distribution of stars in our halo sample in IoM (top panels) and in velocity space (bottom panels), as well as the 91 significant clusters (in colour) identified by the algorithm.
Compared to EDR3, the fraction of stars in clusters is similar ($\sim$ 13\%)
as well as their distribution in these spaces. The most striking difference are the new clusters at low binding energy. 

The purple and blue clusters in Fig.~\ref{fig:IoM_Clusters} with $v_z \sim 0$ and large $v_\phi$ have unexpected, possibly spurious kinematics. Because their stars have $|b| < 10\deg$, exhibit higher than average radial velocity errors and their spectra SNR~$<10$, we suspect also here unreliable radial velocities \citep[see][]{Katz2022}. We thus remove these clusters from our analysis leaving 89 significant clusters.

\begin{figure*}
\centering
    \includegraphics[width=1\textwidth]{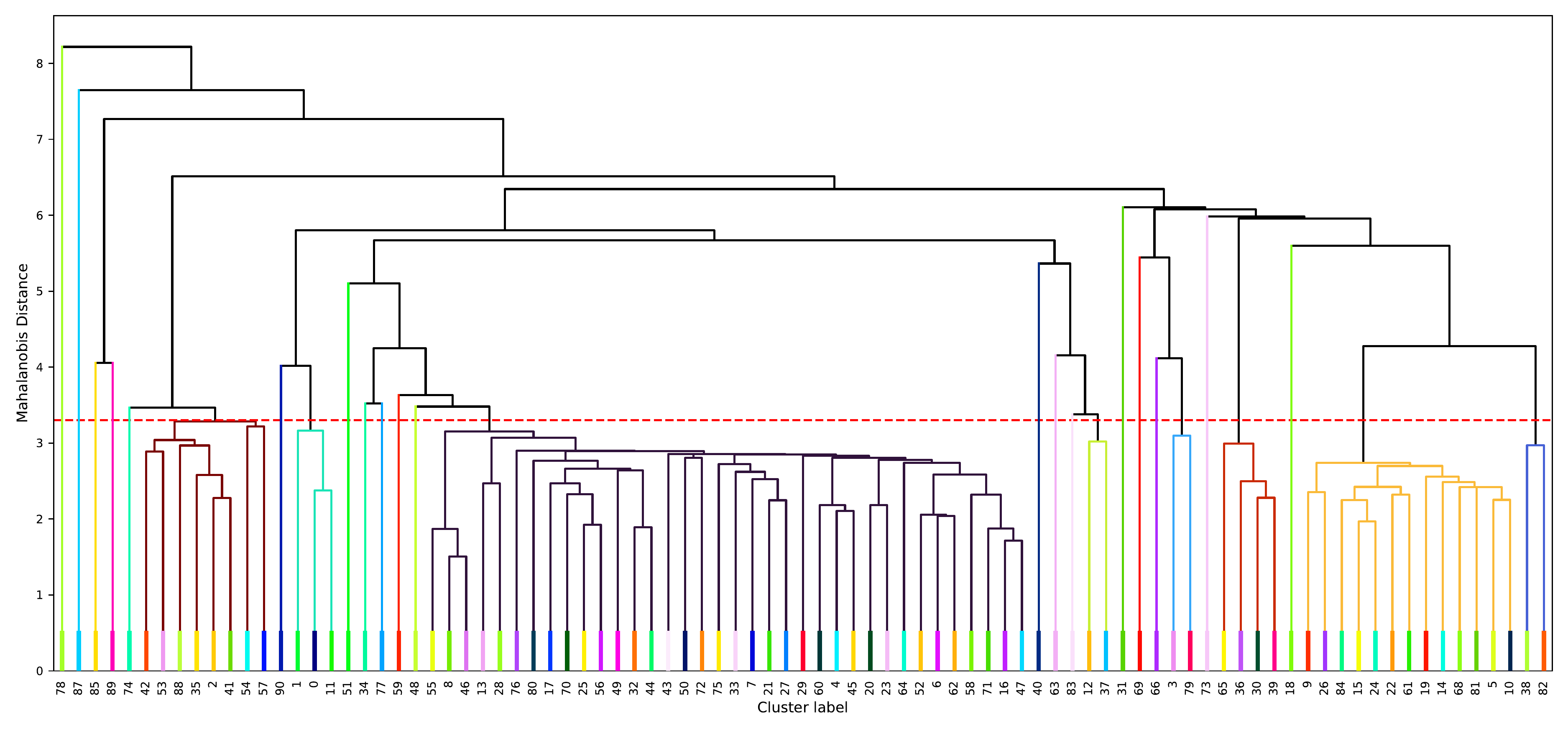}     
    \caption{
    The relationship in IoM space between the significant clusters found,
    shown as a dendrogram using the Mahalanobis distance between the clusters in this space (see main text for details).
    Using this Mahalanobis distance, the clusters are further joined
    up to a cut-off threshold (red dashed line) taken to be 3.3.
    The label of the clusters is given on the $x$-axis, and the clusters that are joined to form larger groups have links of the same colour.
    Using this distance cut, we find 7 preliminary main groups, one cluster pair and 19 individual clusters.
    }
    \label{fig:dendrogram}
\end{figure*}

\begin{figure*}
    \centering
    \includegraphics[width=\linewidth]{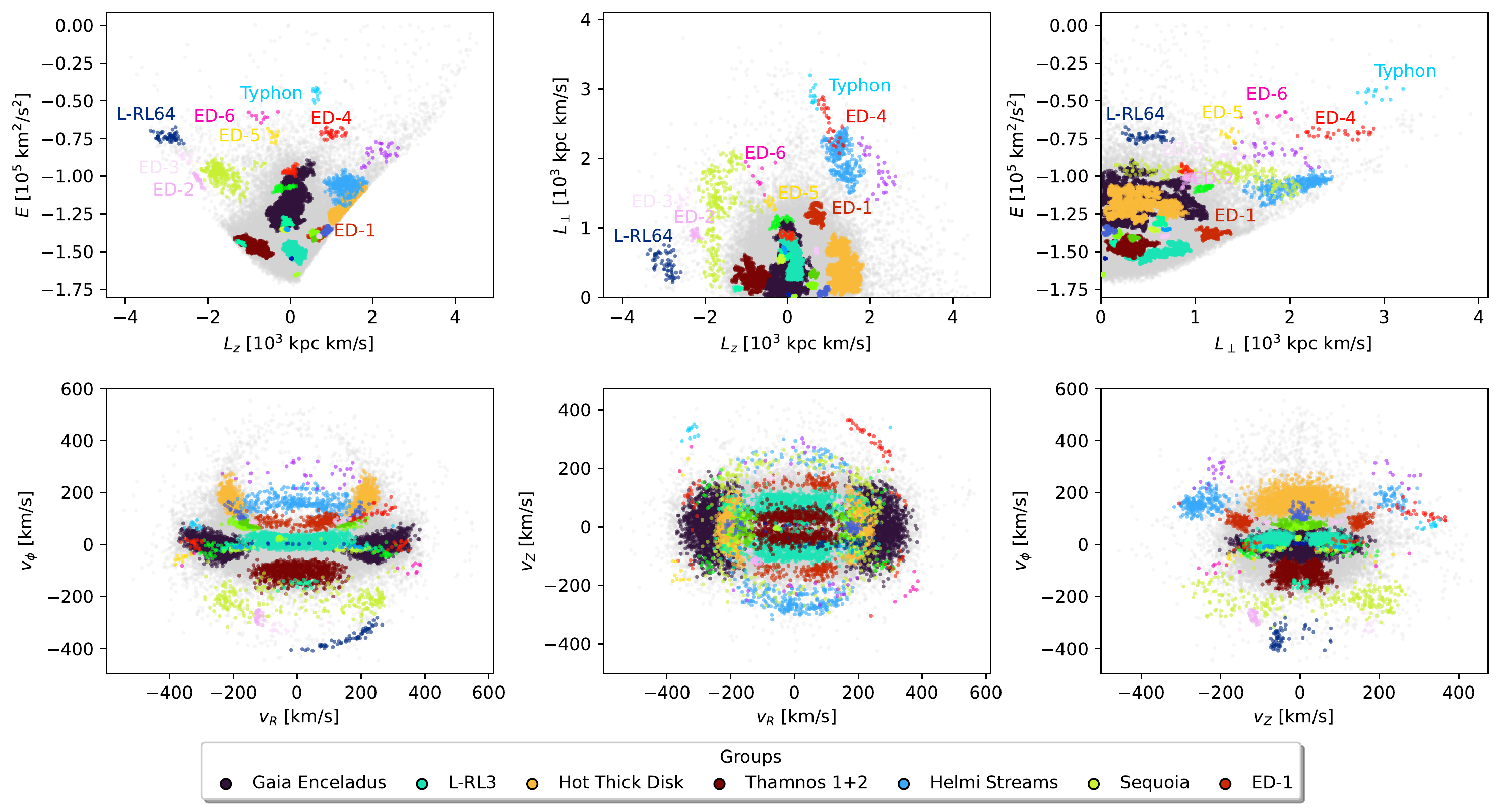}
    \caption{
    The members of our joined groups and individual clusters, where different colours indicate a star's association. Stars that are not part of clusters or groups are shown in the background in grey.
    The  top and bottom rows shows respectively, IoM and velocity space.
    }
    \label{fig:IoM_Groups}
\end{figure*}

Fig.~\ref{fig:dendrogram} shows a dendrogram linking the 89 
clusters by Mahalanobis distance.
Several large groups are formed by clusters linking at small $D^{\prime}$, as seen previously with EDR3 \citep{lovdal2022}. We tentatively set a limit at a Mahalanobis distance ($D^{\prime}_{\textrm{lim}}$) $\sim$~3.3 (red dashed line in Fig.~\ref{fig:dendrogram}).
This Mahalanobis distance is motivated by our analyses of EDR3 
and what we know from the literature on the halo so far \citep{koppelman2019,naidu_evidence_2020}.
Certain regions of IoM space are sensitive to our choice of $D^{\prime}_{\textrm{lim}}$.
For example, imposing a $D^{\prime}_{\textrm{lim}}$ larger than 3.3 linked together Sequoia and what we believe to be a separate smaller cluster which we label as ED-3 in Fig. \ref{fig:IoM_Groups}. 

\subsection{Main Groups}

Our tentative $D^{\prime}_{\textrm{lim}}$ cut suggests we may identify 7 primary groups, one small pair of clusters, and 19 independent clusters. 
The majority of these groups correspond to previously identified substructures. 

To characterise better each of the groups and remaining individual clusters, we proceed to define core members with a Mahalanobis distance in IoM space to each group/cluster $D <2.13$ (this cut corresponds to the value of $D$ containing 80\% of the cluster/group members, and was found to minimise noise when adding tentative members \citealt{lovdal2022,ruizlara2022}). We also add tentative members by identifying all stars within 5 kpc from the Sun and after applying the same quality criteria we adopt the same Mahalanobis distance cut to each group/cluster. 
This results in 31,653 stars (more than 3$\times$ the original number of members) in a group or individual cluster.

\begin{figure*}
    \centering
    \includegraphics[width=0.92\linewidth]{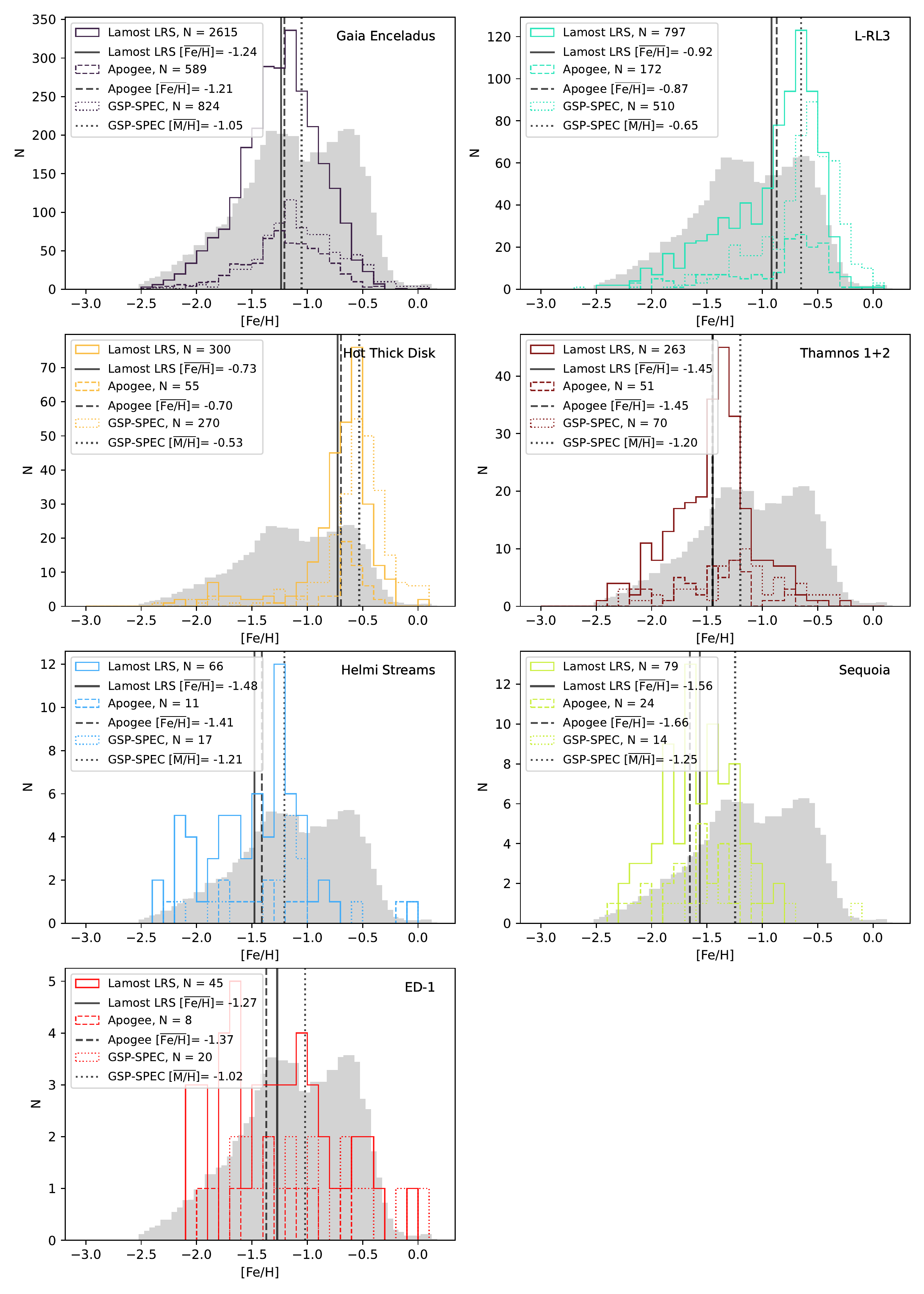}
    \caption{Metallicity distributions for the 7 groups identified in this study that have metallicity information. Solid, dashed and dotted lines correspond to metallicities from LAMOST LRS, APOGEE and GSP-SPEC, respectively. The grey histogram in the background shows the entire halo sample LAMOST LRS metallicities normalised for comparison with each group.
     }
    \label{fig:Groups_MDFs}
\end{figure*}

\begin{figure}
    \centering
    \includegraphics[width=\columnwidth]{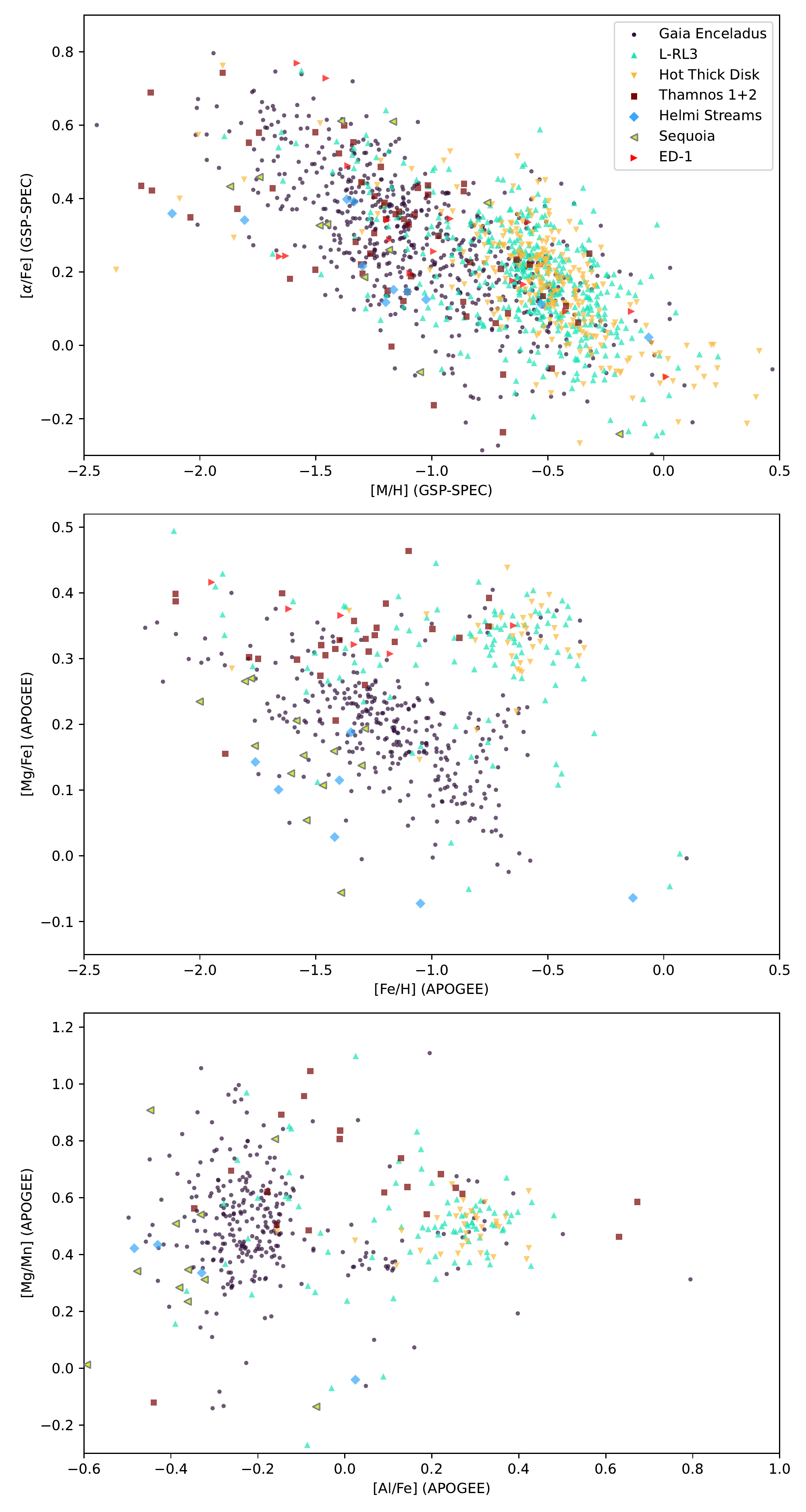}
    \caption{Chemical abundances from GSP-SPEC (top) and APOGEE (bottom panels). The top two panels can be compared as they show [M/H] vs [$\alpha$/Fe] and [Fe/H] vs [Mg/Fe]. The bottom panel showing [Al/Fe] vs [Mg/Mn] is often used to separate accreted from in-situ stars.
    The legend shows groups or individual large clusters identified in our analysis.}
    \label{fig:alpha_fe}
\end{figure}

We now discuss the properties of the different groups and clusters identified. The metallicity distributions (MDF) and their abundance patterns are shown in Figs.~\ref{fig:Groups_MDFs} and \ref{fig:alpha_fe}. The largest number of stars with a metallicity measurement stems from the LAMOST LRS set. It is reassuring to see in Fig.~\ref{fig:Groups_MDFs} that the MDFs obtained using GSP-SPEC and APOGEE are very similar, modulo the smaller number of stars (and possibly a small offset). We also note that the MDFs obtained with original or added members are very consistent with each other. 

The largest group is \textit{Gaia}-Enceladus, with 2872 stars and 36 linked clusters. These stars can be seen to trace the halo peak of the MDF, see panel 1 of Fig.~\ref{fig:Groups_MDFs} and they form a clear sequence in [$\alpha$/Fe] space, see Fig.~\ref{fig:alpha_fe}. Note that GSP-SPEC offers a slightly less clear distinction between the sequence defined by {\it Gaia}-Enceladus stars and the hot thick disk \citep[see also][]{RecioBlanco2022}, which is why we plot separately in the middle panel of Fig.~\ref{fig:alpha_fe}, [Mg/Fe] vs [Fe/H] from
APOGEE. Following \citet{Horta2021}, we also show [Mg/Mn] vs [Al/Fe] based on APOGEE, which is a useful chemical space to separate more clearly accreted from in-situ stars.

The second largest group is shown in cyan in Fig.~\ref{fig:IoM_Groups} at low energy. This group is very similar to the Cluster 3 identified in \citet{lovdal2022,ruizlara2022} and we, therefore, refer to it as L-RL3. It contains 1958 stars and is made up of 3 clusters. The MDF shows that this group is a mix of two populations: a high metallicity population (akin to that of the hot thick-disk) and a well-populated low metallicity tail, see Fig \ref{fig:Groups_MDFs}. This can also be seen in the middle panel of Fig.~\ref{fig:alpha_fe} where the high-metallicity stars in this group populate both high-alpha sequences, while the low metallicity stars seem to define a sequence parallel to that of Gaia-Enceladus but with lower [Mg/Fe]. The mix of in-situ and accreted populations is confirmed from their distribution in [Mg/Mn] vs [Al/Fe] space.

The third-largest group corresponds to the heated (or ``hot") thick disk stars, containing 1450 stars. It is shown in orange in Fig.~\ref{fig:IoM_Groups}, and is made up of 13 clusters, indicating a large amount of substructure (e.g. stripes in energy) in this component. 
Its MDF, the third panel of Fig.~\ref{fig:Groups_MDFs}, shows very little contamination from the metal-poor halo peak. The abundances in Fig.~\ref{fig:alpha_fe} (small orange triangles) show the characteristic high [Mg/Fe] at high metallicity of this in-situ component.

Thamnos 1 and 2 can be seen in brown in Fig.~\ref{fig:IoM_Groups}, composed of 851 stars and 8 clusters. Interestingly some of the stars in this structure appear to define a tight and distinct chemical sequence with very low [$\alpha$/Fe] in the top panel of Fig.~\ref{fig:alpha_fe} (brown squares). The APOGEE abundances suggest a contribution from in-situ stars, but also clearly demonstrate the presence of accreted stars (bottom panel). Although in previous work by \citet{koppelman2019,ruizlara2022}, we identified two sub-components in Thamnos, our preliminary analysis does not warrant (yet) such a separation.

The Helmi streams (shown in light blue in Fig.~\ref{fig:IoM_Groups}) are made up of 319 stars and 2 clusters. In this run both $L_{\perp}$ clumps of the Helmi streams \citep[see ][]{dodd2022} are identified as one single cluster (although each clump is of similar significance) and joined to a smaller cluster within the lower $L_{\perp}$ clump. The few stars with abundance information indicate low values of [$\alpha$/Fe].

Sequoia can be seen in light green in IoM space in Fig.~\ref{fig:IoM_Groups}, consisting of 247 stars and made up of joining 2 clusters. Its MDF shows hints of multiple peaks \citep[see also][]{naidu_evidence_2020}, which do not, however, appear to correspond to separate dynamical structures.
Fig.~\ref{fig:alpha_fe} suggests that these stars (green triangles) follow a distinct chemical sequence from other halo substructures; for a fixed metallicity, they have lower $\alpha$-abundances, as can be seen both from GSP-SPEC and APOGEE data.

One new group (we refer to it as ED-1) corresponds to the red structure seen below the ``hot" thick disk  in energy in Fig.~\ref{fig:IoM_Groups}. It contains 246 stars and is made up of 4 clusters. Its MDF spans a wide range of metallicities and appears to exhibit several metallicity peaks, roughly corresponding to the hot thick disk, \textit{Gaia} Enceladus, and a more metal-poor relatively prominent peak at [Fe/H]$\sim -1.8$. The abundances reveal stars located in both the accreted and in-situ regions of chemical space. 


\subsection{Remaining Individual Clusters}

Of the 19 clusters left ungrouped with our $D^{\prime}_{\textrm{lim}}$ cut, most have only 10 or fewer stars with metallicities (even after adding members within 5 kpc), and therefore we do not show their MDFs.

The tight pair and the three small clusters located between {\it Gaia}-Enceladus and the hot thick disk in $L_z-L_\perp$ 
have metallicities and abundances mostly consistent with being in-situ. 
Five small clusters overlap in IoM space with the region occupied by \textit{Gaia}-Enceladus (see Fig.~\ref{fig:IoM_Groups}), however, the small number of stars with metallicity information make a clear association inconclusive. The lowest energy cluster shown in Fig.~\ref{fig:IoM_Groups}  corresponds the globular cluster M4. 

The most retrograde cluster (in blue in Fig.~\ref{fig:IoM_Groups}, indicated as L-RL64) contains 59 original members and 7 stars have LAMOST LRS metallicities with a mean of $-1.67$.
It has a higher energy than Sequoia, and its kinematics are also clearly distinct.
This cluster has been identified before by \citet[][as cluster 64, see their Fig.~11, and re-discovered by \citealt{Oria_Anaeus_22_2022arXiv220610404O}]{ruizlara2022}, where it was argued to be independent given Sequoia's estimated mass (from its mean metallicity) which would be inconsistent with such a large extent in IoM space \citep{koppelman2019}. The one star with APOGEE abundances have lower [Mg/Fe] than \textit{Gaia}-Enceladus.

There is another smaller cluster located close to Sequoia in IoM space (ED-2, in pink in Fig.~\ref{fig:IoM_Groups}). 
With 32 original member stars, only 3 have a LAMOST LRS metallicity, but they are similar, namely [Fe/H] = $-1.88$, $-2.07$ and $-2.19$. This cluster is extremely tight in velocity space, as can be seen 
in the bottom row of Fig.~\ref{fig:IoM_Groups}, and it 
appears to form a stream in $x-z$ space. 

The last small highly retrograde cluster 
located in this region of IoM space, can be seen in Fig.~\ref{fig:IoM_Groups} in light pink. ED-3 contains 16 original member stars, of which two have LAMOST LRS [Fe/H] of $-1.48$ and $-1.47$,  and one has [Fe/H] = $-1.37$ from APOGEE, also suggesting a rather small spread in metallicity. Intriguingly, the APOGEE star has a low [Mg/Fe], but it is located in the in-situ part of the [Al/Fe] - [Mg/Mn] space.

Two clusters are located in Fig.~\ref{fig:IoM_Groups}
directly above the Helmi streams in energy and $L_\perp$ at similar $L_z$.
The cluster with the lower energy, ED-4, contains 29 original member stars and has 4 stars with a LAMOST LRS metallicity ranging from $< -1.6$ to $-1.05$, and one star with a low [Mg/Fe] abundance in APOGEE, suggesting an accreted origin. 
The cluster just above ED-4 in IoM
(cyan in Fig.~\ref{fig:IoM_Groups}) overlaps with the recently reported Typhon \citep{Tenachi_Typhoon_22_arXiv220610405T}.
It contains 12 stars, 3 of which have a very similar LAMOST LRS metallicity; namely 
$-1.23$,  
$-1.24$ 
and $-1.32$. 
Having three stars with such similar metallicities makes this a very interesting cluster. Unfortunately, we do not have any abundance information for these stars. 
 
Located at high energy and with retrograde motion, ED-5 in yellow in Fig.~\ref{fig:IoM_Groups}, contains 
12 stars. The LAMOST LRS metallicities for three members are $-1.46$, $-1.21$ and $-1.22$, with the latter two stars having the same metallicity in APOGEE. These two APOGEE stars are on the low $\alpha$ track and both fall within the accreted region of the [Al/Fe] - [Mg/Mn] space. Therefore, this is a potentially interesting cluster to follow up further.

The cluster above in IoM space we call ED-6 (shown in pink in Fig.~\ref{fig:IoM_Groups}) and it contains 10 original members, 6 have a LAMOST LRS metallicity and show a very small spread around -1.3, except for one star at -0.86. 
One overlapping star has an APOGEE abundance placing it in the boundary of accreted vs in-situ in [Al/Fe] - [Mg/Mn] space but this is the outlier star with [Fe/H] of -0.86 dex. 

There are three remaining clusters, one at low energy overlapping with L-RL3 in IoM space, one overlaps with Thamnos and the other cluster close to the Helmi streams in $L_{\perp}$. None of these have sufficient metallicity information to be able to comment further. 

In summary, in this section we have identified 7 main groups, and of the preliminary 19 individual clusters 8 of which can be tentatively associated to the larger groups. 


\section{Discussion and conclusions}\label{sec:conclusion}

We have constructed a sample of dynamically selected nearby halo stars based on the {\it Gaia} DR3 dataset. Using a single-linkage based algorithm, and thanks to the excellent quality of this dataset, we have identified 89 clusters in Integrals of Motion space. By grouping these according to their Mahalanobis distance in this space, and subsequently by comparing the metallicities of the member stars using data from GSP-SPEC, APOGEE and LAMOST LRS, we have been able to identify 7 groups and 11 individual independent clusters. Out of the 7 large groups, 6 have already been reported in the literature, namely {\it Gaia}-Enceladus, the hot thick disk, Thamnos, Sequoia, Helmi streams and L-RL3 \citep[see e.g.][]{ruizlara2022}. ED-1 is a new dynamical group and it contains a mix of populations, and probably includes contamination from the hot thick disk and from Gaia-Enceladus, but its MDF reveals a peak at low metallicity, and the abundances of some of its stars suggest an accreted origin. 

Of the 11 remaining independent clusters, two have been reported before: one is the globular cluster M4, and the other one is L-RL64 \citep{ruizlara2022}. A third cluster was reported as Typhon \citep{Tenachi_Typhoon_22_arXiv220610405T} while this manuscript was written. Three clusters do not have sufficient metallicity or abundance information to comment.
The remaining 5 clusters (ED-2, -3, -4, -5 and -6) are all interesting in different ways: most are rather tight dynamically (especially ED-2), and some show a very small spread in metallicities (ED-2, ED-3, tentatively ED-5 and ED-6),
while all of them appear to have been accreted based on their location in IoM space and
the chemical abundances of a few member stars.

Given the complexity of the debris from large accretion events \citep[see e.g.][]{Koppelman2020,amarante2022} some of these smaller clusters may ultimately be related to one another or the larger groups.
To make more progress towards our goal of inferring the assembly history of the Milky Way, we need to probe beyond the immediate solar vicinity, and especially obtain metallicities and more precise chemical abundances for larger numbers of stars.

\begin{acknowledgements}

We acknowledge financial support from a Spinoza prize to AH. 
This work has made use of data from the European Space Agency (ESA) mission {\it Gaia} (\url{https://www.cosmos.esa.int/gaia}), processed by the {\it Gaia} Data Processing and Analysis Consortium (DPAC, \url{https://www.cosmos.esa.int/web/gaia/dpac/consortium}). Funding for the DPAC has been provided by national institutions, in particular, the institutions participating in the {\it Gaia} Multilateral Agreement.

The analysis has benefited from the use of the following packages: vaex \citep{breddels2018}, AGAMA \citep{vasiliev2019}, NumPy \citep{van2011}, matplotlib \citep{hunter2007} and jupyter notebooks \citep{kluyver2016}.

Our catalogue and substructures will be made available online upon acceptance or before upon reasonable request.
\end{acknowledgements}

\bibliographystyle{aa} 
\bibliography{references}

\end{document}